\documentclass[pre,groupedaddress,showpacs,preprint]{revtex4}
\usepackage[english]{babel}
\usepackage{epsfig}
\usepackage{amsmath}
\usepackage{amssymb}
\usepackage{amsfonts}
\usepackage{array}
\usepackage{subfigure}
\usepackage{psfrag}

\begin{document}
\bibliographystyle{apsrev}

\title{Internal friction and mode relaxation in a simple chain model}

\author{S. Fugmann and I. M. Sokolov}

\affiliation{Institut f\"ur Physik,
Humboldt-Universit\"at zu Berlin,
Newtonstra\ss e 15, D-12489 Berlin, Germany}

\date{\today}

\begin{abstract}
We consider equilibrium relaxation properties of the end-to-end distance and of principal components
in a one-dimensional polymer chain model with nonlinear interaction between the beads. While for the single-well
potentials these properties are similar to the ones of a Rouse chain, for the double-well interaction potentials, 
modeling internal friction, they differ vastly from the ones of the harmonic chain at intermediate times and intermediate temperatures. This minimal description within a one-dimensional model mimics the relaxation properties found in much more complex polymer systems. Thus, the relaxation time of 
the end-to-end distance may grow by orders of magnitude at intermediate temperatures. The principal components 
(whose directions are shown to coincide with the normal modes of the harmonic chain, whatever interaction potential is assumed) not only display larger relaxation times but also subdiffusive scaling.
\end{abstract}


\maketitle

\section{Introduction}

The dynamics of polymers and peptides attracted large attention in the past decade, 
mostly due to the relations of such dynamics to biological function. 
First such interest was due to the dynamics of nonequilibrium states
(connected with the folding problem and with the biological functioning of the
proteins), and only later some interest to \textit{equilibrium} fluctuations
in proteins arose. This interest was caused by the two reasons: one
has to do with the thermal stability of proteins as 
connected to their structures \cite{Granek05}, another one is related 
to the luminescent measurements of fluctuations of distance between 
the two groups in equilibrium \cite{Xie04} and the discovery of the anomalous kinetics and of
extremely large characteristics times in such fluctuations. 
The results of these investigations, put together, lead to an enigma: On one hand, 
the thermal stability and many other properties of proteins 
can be well-described within simple bead-spring models of such systems \cite{Togashi07_PNAS,Cressmann08_PRE}, which for small deviations from equilibrium 
can be reduced to a standard picture of normal modes in a complex harmonic network.
The evaluation of the characteristic times in such systems, however, leads to results being off by orders of magnitudes when compared to the observed times \cite{Tang06_PRE}. The strong discrepancy in correlation times implies the existence 
of a strong additional {\it internal} friction mechanism slowing down the dynamics compared the linear Rouse one.

Recently, the anomalous kinetics (power-law decay of correlations) was observed 
even in single modes and even in relatively short and flexible
peptides, i.e. linear chains lacking secondary structure \cite{Neusius08}, and the behavior found here is very close to the one observed in protein simulations \cite{Senet08}. Therefore such kinetics 
probably is an intrinsic property of a large class of polymers, and is not 
connected to the specific properties of the secondary structure of proteins. Furthermore it has been shown that trap models cannot account for the anomalous dynamics \cite{Neusius08}.
Therefore, it is necessary to analyze, what kind of the minimal assumptions
about the intramolecular potentials have to be made to build a model
mimicking the behavior observed in realistic molecular dynamics simulations. 

Typical mechanisms of internal friction involve the existence of barriers in the free energy landscape of the system, being of entropic or of enthalpic nature. The necessity to overcome such barriers slows down the overall dynamics by the Arrhenius factor which might be quite large \cite{Fixman78}. However, the existence of a complex, nonlinear energy landscape may lead to a strong mixing of modes appearing in the linearized description, and make the whole analysis based on such a picture problematic. As we proceed to show, this is not the case: although the dynamics of modes is strongly influenced, the directions of 
the principal components (PCs) in configuration space follow those of the normal modes, which resolves the enigma.

In what follows we first consider the Brownian dynamics of a three-dimensional polymer chain, as applied e.g. for simulating polyethylene molecule within the valent-angle model \cite{Ryckaert75,Rigby87,Binder97}, and show that it follows slow kinetics at intermediate times, provided that the temperature is low enough. On the other hand, at high temperatures the typical Rouse dynamics sets on.

To simplify the model even more, we consider a one-dimensional chain of beads, similar to the
Rouse model of the polymer dynamics, with the only difference being changing the harmonic
interaction between the beads for a nonlinear one. We discuss the end-to-end distance of the chain and the relaxation properties of the PCs \cite{Kitao99} of the system. 

The one-dimensional model discussed here is a close continuous analogue of the so-called ``necklace model" for reptation \cite{Guidoni03,Drz06}, so that the results might have a broader applicability also outside of the scope of present investigations.

The results of extensive numerical simulations show that while a whatever single-well potential tested 
does not affect drastically single-mode relaxation properties, the double-well potentials lead to 
intermediate-time behavior quite similar to the one observed in realistic systems and to 
strong increase in relaxation times compared to the harmonic case. Therefore such a model
may be considered as a possible candidate for the explanation
of the corresponding findings. 
The multiwell intramolecular potentials appear at different scales and 
are ubiquitous in polymers \cite{Ryckaert75,Binder97}. They appear quite naturally e.g. within the valent-angle model.
The anomalous behavior is present within the finite 
temperature range (which in the realistic case may be relevant for the biological functioning), 
and disappears both for low and for high temperatures, where the dynamics can therefore be described
by an effective harmonic model, albeit with the values of parameters strongly different from the 
``microscopic'' ones. This property can explain that the generalized Gaussian models 
work quite well in predicting thermal stability properties while failing to 
mimic the temporal fluctuation behavior at moderate temperatures. 

The work is organized as follows: In the next section we present the relaxation properties found in a complex three-dimensional polymer model and introduce in Sec.~\ref{s_model} the one-dimensional chain model. In Sec.~\ref{s_relate} we consider the relaxation properties of the end-to-end distance of the chain and discuss in Sec.~\ref{s_pcs} the behavior of its PCs. We focus on their directions and their temporal scaling, which is found to coincide with the one found in the realistic three-dimensional model. Finally we summarize our results.

\section{Internal friction in a three-dimensional chain model}
\label{s_comp}

In the first part of the work we consider the relaxation properties of a complex polymer model which was proposed to mimic a polyethylene molecule. The model equations can be found in \cite{Ryckaert75,Rigby87,Binder97}. They include valence bond-, valence angle- and torsional angle-interactions. For our purpose we neglect the Lenard Jones interactions and focus on the influence of the angle interactions which follow a multistable potential energy landscape. All parameter values are the same as in \cite{Rigby87} except for the bond length $\tilde{l}_0=1.$, and the constants $k_b=3.$, $k_{\Theta}=3.$, and $k_{\phi}=0.1$ which only set the timescale of simulation. We perform Brownian dynamics simulations and study the relaxation properties of the PCs and their autocorrelation functions. 

For a discrete set of $M$ measurements of a trajectory $x(t)$ the unbiased autocorrelation function (ACF) of 
an observable $x(t)$ (coordinate, end-to-end distance, etc.) $\phi(t)$ ($t\in 0,...,M-1$) is given by
\begin{equation}
\phi(t)=\frac{1}{(M-t)\sigma^2}\sum_{n=0}^{M-t-1}\left(x(n+t)-\mu\right)\left(x(n)-\mu\right)\,,
\label{eq:ACF}
\end{equation}
with the sample mean $\mu$ and sample variance $\sigma^2$. Since (for $\mu=0$) the mean squared displacement
$\langle x^2(t)\rangle=\langle \left(x(t+t')-x(t')\right)^2\rangle
=\langle x^2(t+t')-2x(t+t')x(t')+x^2(t')\rangle
=2\sigma^2\left(1-\phi(t)\right)$, the values of 
$1-\phi(t)$ and $\langle x^2(t)\rangle$ (e.g. used in \cite{Neusius08}) contain the same information and differ only in normalization. We denote the unbiased ACF of the unbiased ACF of the $k-$th PC by $\phi_{k}$. 

\begin{figure}
\begin{center}
\subfigure[]
{\includegraphics[height=6.5cm,width=7.5cm]{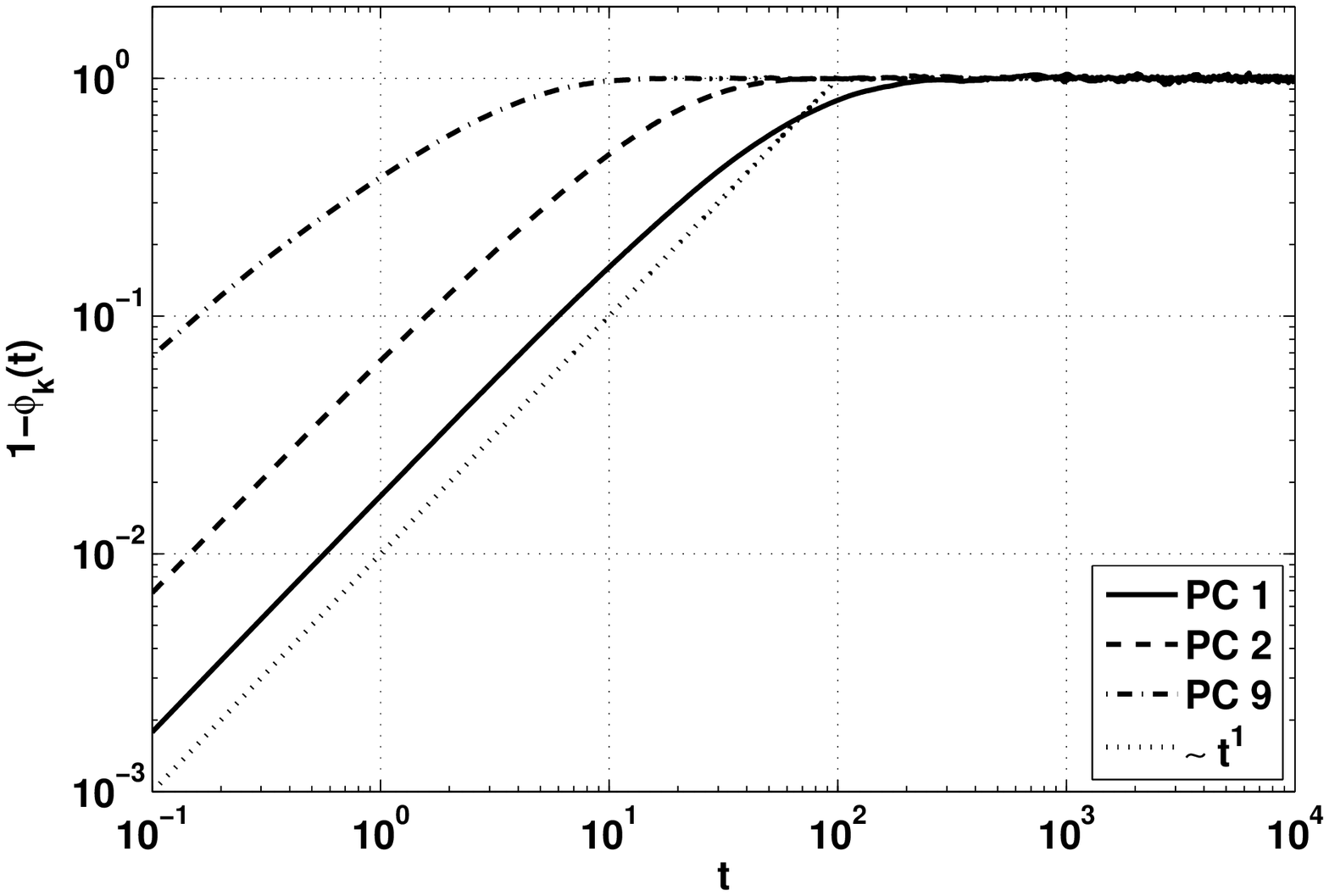}}
\subfigure[]
{\includegraphics[height=6.5cm,width=7.5cm]{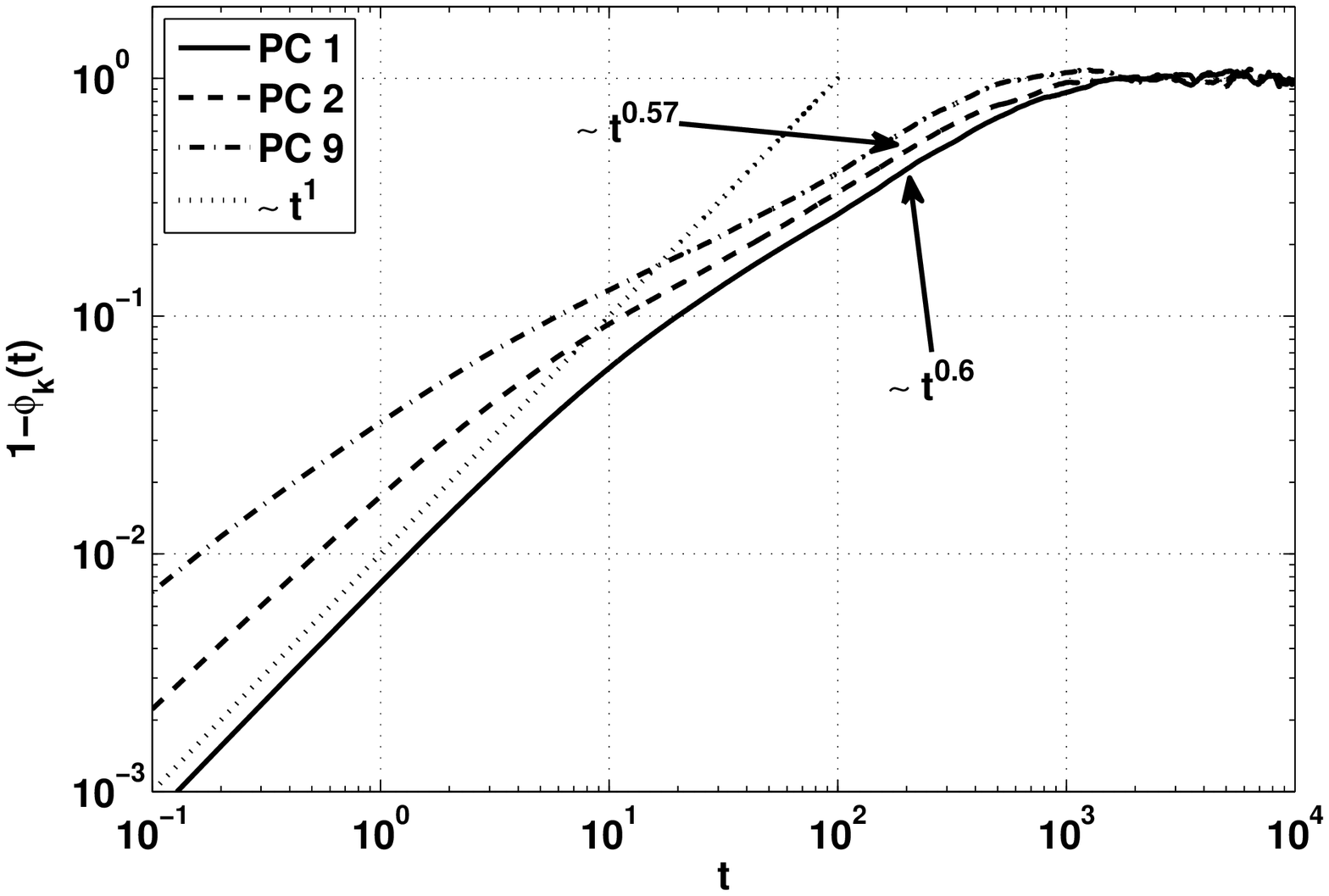}}
\caption{\label{fig:CMPCs}}
\end{center}
\end{figure}

In Fig.~\ref{fig:CMPCs} we show the scaling of $1-\phi_k(t)$ for some of the PCs. In panel (a) we have chosen a high value of the temperature yielding a trans-gauche barrier height of $\Delta E_{tg}/k_BT=0.66$, a cis barrier height of $\Delta E_{c}/k_BT=2.4$ and $\Delta E_{t\rightarrow g}/k_BT=0.16$. All PCs show a scaling $t^{\alpha}$ with $\alpha=1$, which is the same as for the relaxation of single modes in the Rouse chain \cite{Doi96}. The scaling for a lower value of the temperature with $\Delta E_{tg}/k_BT=7.6$, a cis barrier height of $\Delta E_{c}/k_BT=27.6$ and $\Delta E_{t\rightarrow g}/k_BT=1.8$ is presented in panel (b). The relaxation behavior strongly differs compared to the previous case. For small times $1-\phi_k(t)$ still follows follows the $t^{1}$ scaling (dotted line); but at longer time it crosses over to anomalous, subdiffusive behavior. The curve's slope is slightly higher for the first mode. Compared to the high temperature case, the typical relaxation times are shifted to longer times. Thus the existence of barriers in the free energy landscape at intermediate temperatures slows down the dynamics. However, the theoretical analysis of a three-dimensional model is still too involved, so that we simplify it even more in the next section.

\section{The minimal model}
\label{s_model}

In order to mimic the behavior found in the complex three-dimensional model we consider a one dimensional chain of $N$ beads with coordinates $q_1$, ..., $q_N$. The interaction potential is given by $W(l_i)$, with $l_{i} = q_{i}-q_{i-1}$ being the relative displacements of the neighboring beads, $i=1,\cdots,N$. The chain is free, hence $W(l_0)=W(l_{N+1})=0$. For a harmonic potential $W(l_i)$ this model corresponds to a Rouse chain in 1d. The dynamics is overdamped and the system of coupled Langevin equations reads
\begin{equation}
\dot{q}_i=-\frac{1}{\gamma}\frac{d}{dq_i}\left\{W(l_{i+1})+W(l_i)\right\}+\sqrt{2\frac{k_BT}{\gamma}}\xi_i\,,
\label{eq:langevin}
\end{equation}
with $\xi_i$ being Gaussian white noise of strength $k_BT/\gamma$. The friction coefficient $\gamma$ is set to unity in what follows. The simulations correspond to solving the set of coupled equations \eqref{eq:langevin} by use of a Heun integration scheme. In order to get reliable statistics numerical simulations have to cover many orders of magnitude in time. 
Therefore we confine ourselves to relatively short chains consisting of $N=25$ beads. 

We concentrate on three simple prototypes of coupling potentials $W(l_i)$, either single-well or double-well. For the single-well coupling potentials we consider a soft and asymmetric Toda interaction (T-potential) \cite{Toda89} and a hard and symmetric quartic interaction (Q-potential), i.e, $W(l_i)=l_i+\exp(-l_i)-1$ and $W(l_i)=\frac{1}{2}l_i^2+\frac{1}{4}l_i^4$, respectively.
The double-well potential (DW-potential) has the form $W(l_i)=\frac{a}{4} l_i^4-\frac{b}{2} l_i^2$. The latter potential exhibits two minima separated by a maximum of height $\Delta E=b^2/(4a)$. We define $e=\Delta E/(k_BT)$. 
Note, that for small deviations from their equilibria all potentials can be approximated by a harmonic spring with coupling constant $\kappa=1$ for the T-and Q-potentials and $\kappa=2b$ for the DW-potential.

\section{Relaxation of the end-to-end distance}
\label{s_relate}

\begin{figure}
\begin{center}
\subfigure[]
{\includegraphics[height=6.5cm,width=7.5cm]{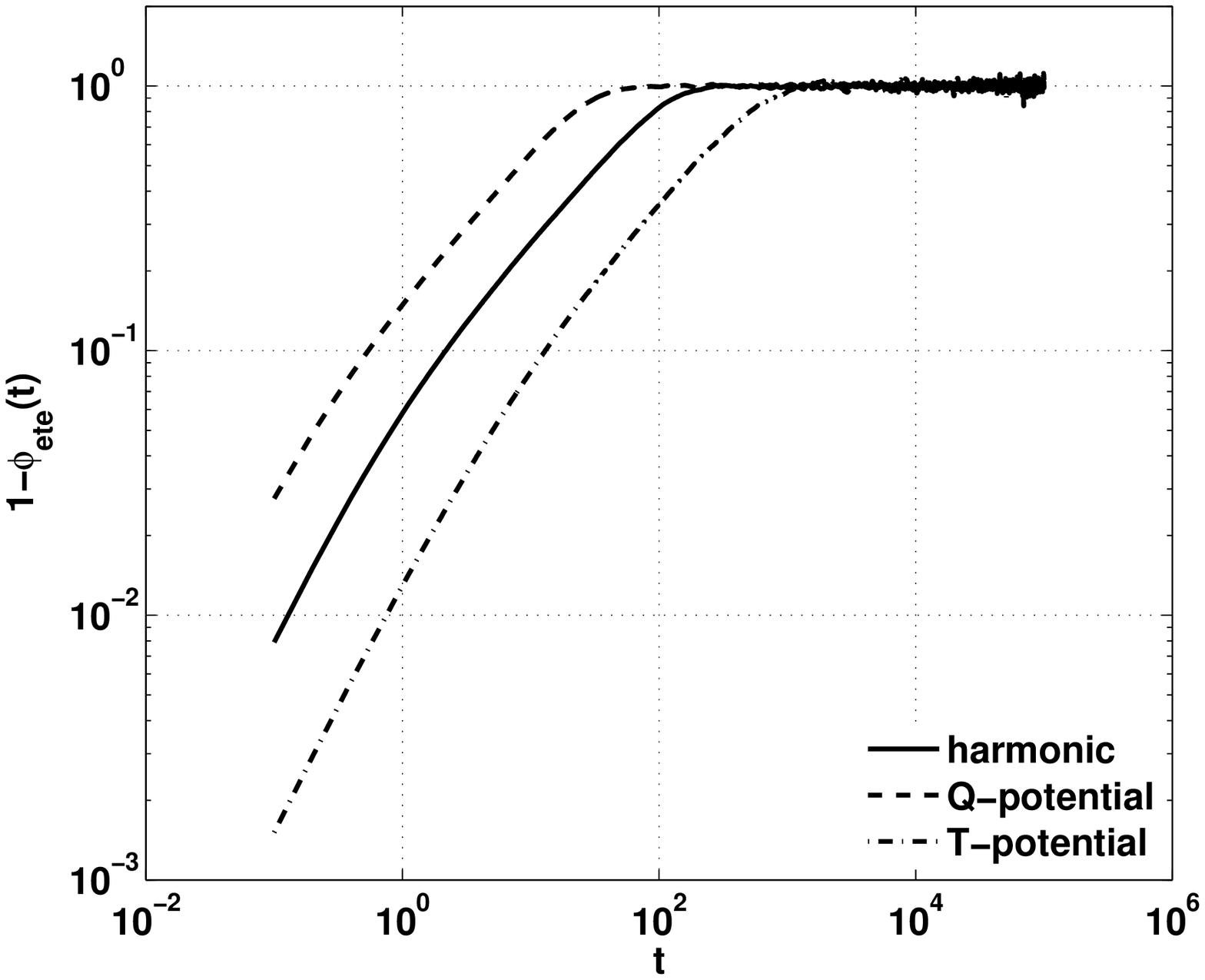}}
\subfigure[]
{\includegraphics[height=6.5cm,width=7.5cm]{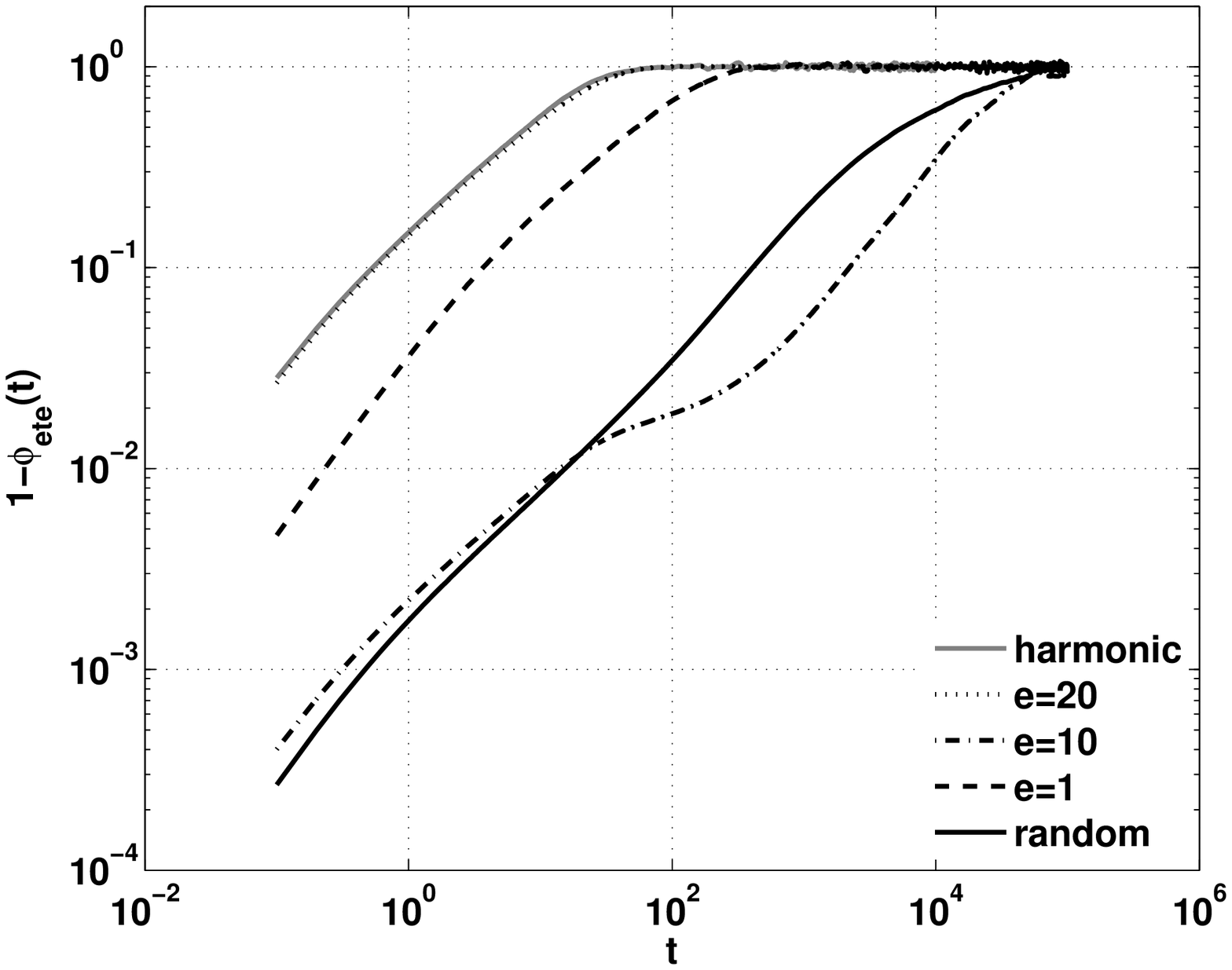}}
\caption{\label{fig:dwETE}}
\end{center}
\end{figure}

First, let us consider the ACF of the end-to-end distance $R_{ete}=q_N-q_1$, see Fig.~\ref{fig:dwETE}. We denote the unbiased ACF of the end-to-end distance by $\phi_{ete}$. Panel (a) shows the scaling of $1-\phi_{ete}$ vs. time for the different single-well potentials. Compared to the corresponding harmonic chain, the autocorrelation time reduces for the Q-potential while it grows for the T-potential, and this effect is stronger for higher temperatures.  An explanation is as follows:  the relaxation time of $R_{ete}$ is to a good approximation the largest relaxation time of the harmonic chain. This one behaves as $\tau_1\sim 1/\kappa$ \cite{Doi96}. For a hard Q-potential the effective value of $\kappa$ is larger than for a harmonic one and grows with temperature, so that $\tau_1$ is expected to be smaller, while for the soft T-potential the opposite is true. In panel (b) $1-\phi_{ete}$ is depicted for the DW-potential. For different values of $e$ the curves strongly differ. At a low temperature ($e=20$, dotted line) the curve coincides with the one of the corresponding harmonic chain (solid gray line) for a finite simulation time (the beads hardly jump between the two minima of the coupling potential): this is exactly what should be observed when such a conformation change mechanism is frozen out \cite{Tournier03_PRL}. For intermediate barrier heights ($e=10$, dashed-dotted line) the behavior of the ACF is strongly different. 
It shows distinctly different behaviors for short and long times interpolated by a plateau. The characteristic relaxation time in the long-time domain (where the kinetics is dominated by  rare fluctuations following the Arrhenius law) is by about four orders of magnitude larger than for the harmonic chain.
We also considered a case when the barrier heights are randomly distributed with the density $\rho(e)\sim \exp(\lambda (e-e_{min}))$, $e\geq e_{min}=5$, $\lambda=0.1$ with a cut-off at $e_{max}=11$ to avoid freezing (solid black line). In this case the plateau is smoothed out, 
but the increase of characteristic relaxation time persists.  
For high temperatures ($e=1$, dashed line) the correlation time is much smaller compared to the previous case (but still larger than in the harmonic case due to the softer potential) and its scaling follows the one of the harmonic chain. We conclude that the increase of the correlation time (related to the timescale of rare barrier crossings) becomes large at intermediate temperatures. Qualitatively the observed behavior is in full agreement with the one made for the realistic three-dimensional model. 

\section{Principal components (PCs)}
\label{s_pcs}

\subsection{Directions of the PCs}

Let us now turn to the behavior of principal components. We proceed to show that in a 
homogeneous linear chain with symmetric interaction potentials
$W(l_i)$ between the beads, with the overall potential energy given by
$\tilde{W}=\sum_i W(l_{i})$, the PCs
follow the directions of the normal modes of the harmonic chain,
independently on the exact form of the potential $W(l_i)$. This is due to
the fact that the directions of PCs are essentially thermodynamical quantities.

The proof of this fact involves the following steps. Consider a
grafted chain of $N+1$ beads, 
whose first bead is attached to the origin of
coordinates by a weak spring, with the small elastic constant
$\varepsilon$ (an ``asymptotically free'' chain). Let $\mathbf{q}$ be the vector
of the coordinates, and $\mathbf{l}$ be the vector with components $l_0=q_0$ and  
$l_i$ for $i \geq 1$. For a given interaction potential the distribution
of $\mathbf{l}$ is given by
\begin{equation}
p(l_0,...,l_n)=\frac{1}{Z} \exp \left[-\frac{1}{k_BT}\left(
\frac{\varepsilon}{2}l_0^2 +\sum_i W(l_{i})  \right)\right].
\label{Canonical}
\end{equation}
This overall canonical distribution factorizes in the product of the distributions
of single $l_i$-components, which means that the corresponding random variables
are independent. Due to symmetry of all interaction potentials, the mean values of
$l_i$ are zero. Hence, the corresponding variables are uncorrelated too: 
$\left\langle l_0^2 \right\rangle =l_0^2 = 2k_BT/\varepsilon$, 
$\left\langle l_i^2 \right\rangle =  l^2$ for $i=1,...,N$ and 
$\left\langle l_i l_j \right\rangle = 0$ for all $i \neq j$. 
The variables $q_i$ are obtained through $l_i$ via linear transformation, 
$q_i = \sum_{j=0}^i l_j$, so that their correlator $C_{ij} = 
\left\langle q_i q_j \right\rangle = l_0^2 
+ \min\{i,j\} l^2 $. The matrix $\hat{C}$ can be diagonalized. Its eigenvectors
are the PCs of the grafted chain, whether harmonic or not. 
The normalized PCs can not depend on $l^2$ or $l_0^2$ independently, but only
on their ratio $l^2/l_0^2 = l^2 \varepsilon/(2k_BT)$. Note that for the free
chain, $\varepsilon \to 0$, $l_0 \to \infty$, the PCs of a harmonic and anharmonic
chain coincide. The same is true also for asymmetric interaction potentials, where the lengths $l_i$ 
have to be corrected for thermal expansion.  
 
Turning to a harmonic chain, we can change $l_i$ for $q_i$ in Eq.(1), which
now reads 
\begin{equation}
p(q_0,...,q_N)=\frac{1}{Z'} \exp \left[-\frac{\kappa}{2k_BT}\left( \mathbf{q} \hat{M} \mathbf{q} 
\right)\right],
\label{Canonical2}
\end{equation}
where $\hat{M}$ is the tridiagonal force matrix, whose elements are:
$m_{00} =1+\varepsilon$, $m_{NN}=1$, the rest of diagonal elements
$m_{ii}=2$, the elements on the sub- and super-diagonals are equal to $-1$.
Since the corresponding distribution is a multivariate Gaussian, the elements of
the correlation matrix $\hat{C}$ are proportional to the ones of the 
inverse of the matrix $\hat{M}$.
The eigenvectors of the matrix $\hat{M}$ are the normal modes of the harmonic chain.
Since the matrix $\hat{M}$ and its inverse $\hat{C}$ shear the eigenvectors,
those are also the PCs of the harmonic chain.   The idea of considering the grafted chain 
is connected to the wish to have $\hat{M}$ invertible.  
In the last step one shows that for  $\varepsilon \to 0$  the limits
of all corresponding eigenvectors exist (the fact having to do with the nondegenerate
nature of a harmonic chain's spectrum). 

Thus assuming whatever interaction potential between neighboring beads, the PCs
follow the directions of the normal modes of the harmonic chain. However, as we proceed to show, 
the dynamical properties of each single mode can be drastically different from the ones of a harmonic chain.

\subsection{Relaxation properties of the PCs}

\begin{figure}
\begin{center}
\subfigure[]
{\includegraphics[height=6.5cm,width=7.5cm]{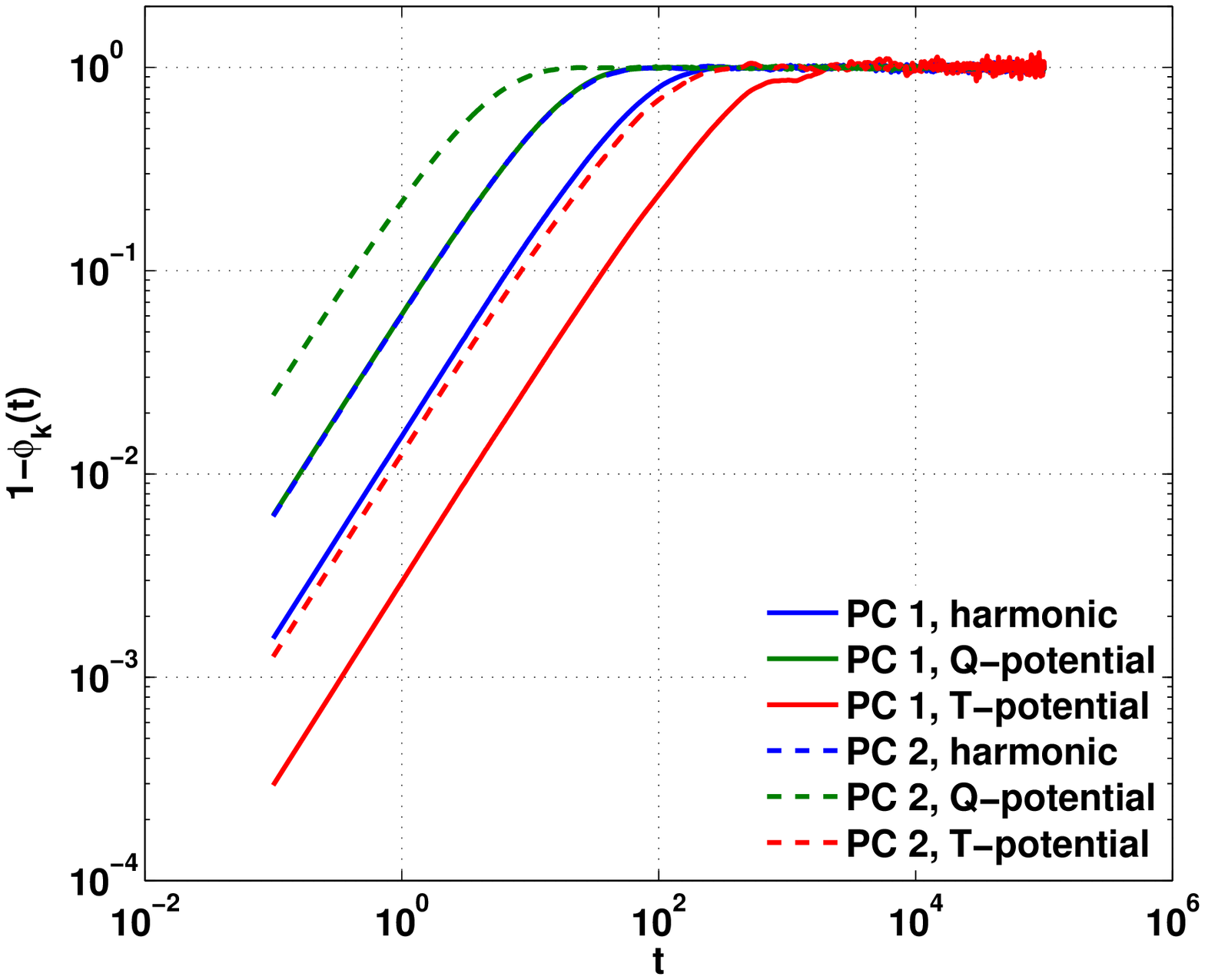}}
\subfigure[]
{\includegraphics[height=6.5cm,width=7.5cm]{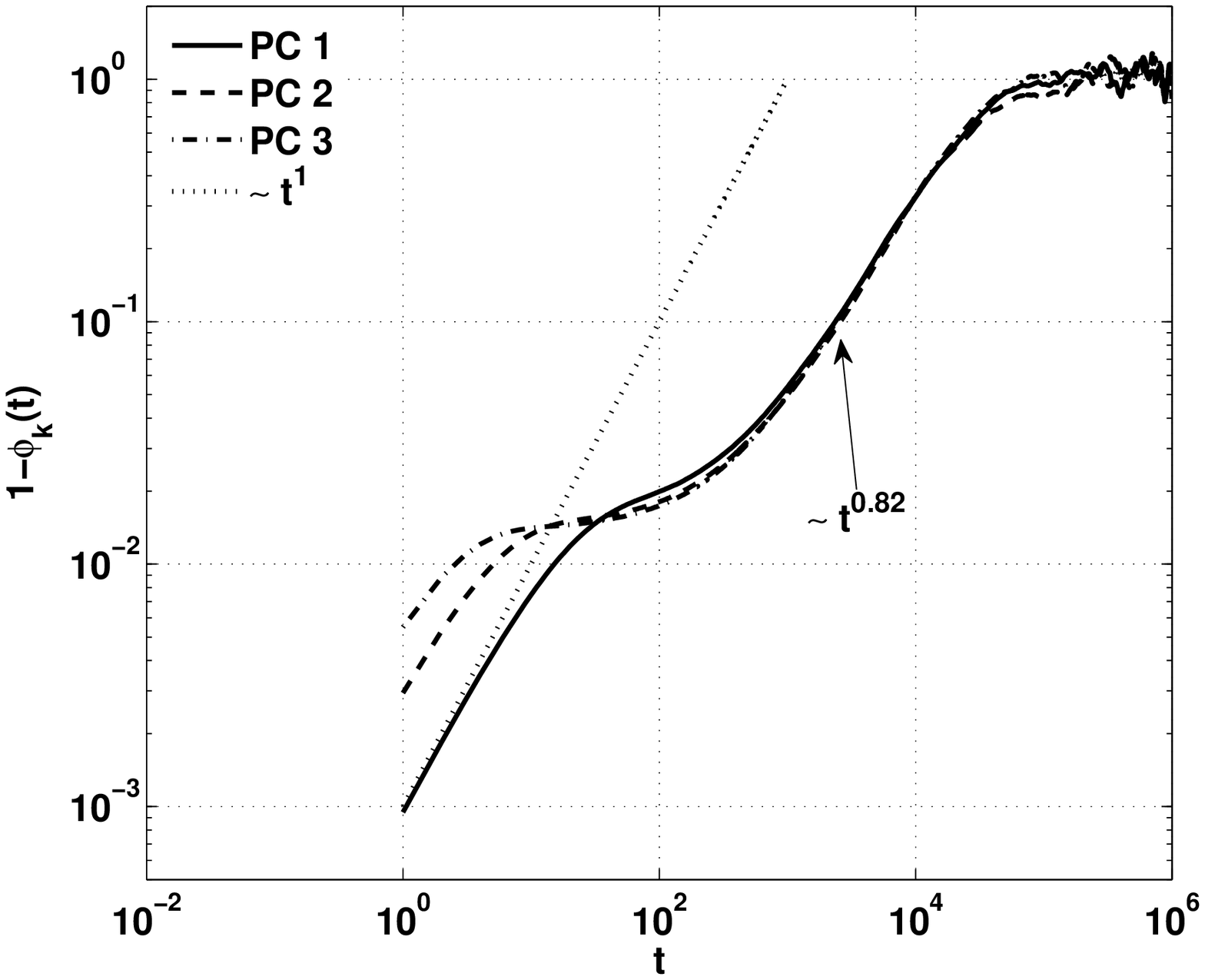}}
\subfigure[]
{\includegraphics[height=6.5cm,width=7.5cm]{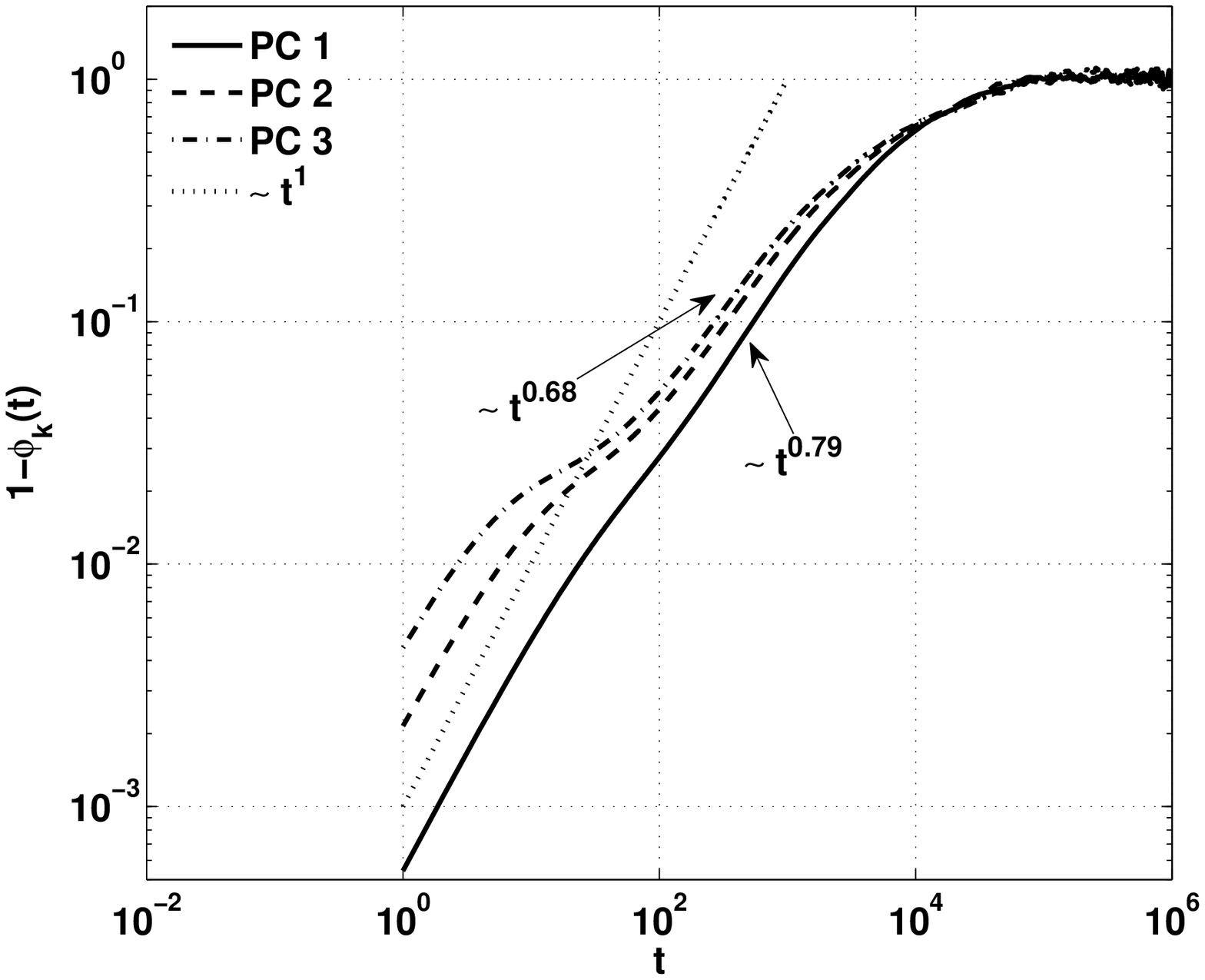}}
\caption{\label{fig:DWPCs}}
\end{center}
\end{figure}

From the normal mode analysis of a harmonic chain it is known that the autocorrelation time of the $k-$th normal mode scales as $\tau_k\sim N^2/k^2$ \cite{Doi96,deGennes79} and does not depend on the noise strength $k_BT/\gamma$. Although the autocorrelation time becomes temperature dependent for the single-well nonlinear potentials, the scaling of $1-\phi_k(t)$ remains the same as in the harmonic situation. This is shown in Fig.~\ref{fig:DWPCs} in panel (a). The curves order relative to the corresponding harmonic limiting case is the same as in Fig.~\ref{fig:dwETE}. Furthermore we find that the scaling with $k$ and $N$ persits (not shown). Thus as in the harmonic chain higher modes have smaller correlation times.

In contrast, for the DW-potential with intermediate barrier height of $e=10$, Fig.~\ref{fig:DWPCs} panel (b), the relaxation times of the lowest three modes are almost equal: the corresponding curves merge for $t>10^2$. The overall shape of the curves is similar to the one for $R_{ete}$, albeit with different scaling. For a harmonic chain the scaling is given by $t^{\alpha}$ with $\alpha=1$ (corresponding to the dotted line in the plot). For small times $1-\phi_k(t)$ still follows follows this scaling; but at longer time it crosses over to anomalous, subdiffusive behavior with $\alpha\approx 0.82$ over two orders of magnitude in time. The case of distributed barrier heights is shown in panel (c). Here we observe again a subdiffusive scaling of the ACF. Interestingly the shown curves for different values of $k$ do not merge at longer times and their slope is mode dependent with smaller $\alpha$ for higher values of $k$ ($k=1$: $\alpha\approx 0.79$, $k=2$: $\alpha\approx 0.72$, $k=3$: $\alpha\approx 0.68$). For modes $k\geq 4$ the slope changes only slightly (not shown). The relaxation behavior of the realistic chain presented in Fig.~\ref{fig:CMPCs} is the same, albeit with somewhat different exponents. However, the exponents are expected to depend crucially on the choice of parameters and the ratio of different timescales in the system. The proposed minimal model is able to mimic qualitatively the relaxation behavior of a much more complex structure and therewith it helps to understand the underlying mechanism of internal friction.

\section{Summary}
\label{s_summary}

To summarize, it was found that under nonlinear interaction potentials, the scaling of the end-to-end ACF is essentially the one of the harmonic chain, though with shifted correlation times. While these changes are not too large for the single-well potentials, for the double-well ones (describing the internal friction) these can lead to the increase in the relaxation times by the orders of magnitude. Turning to the principal components we have shown that for homogeneous chains they follow the normal modes of the harmonic chain, but can exhibit vastly different kinetics, both with respect to the characteristic times and to the overall scaling. For the double-well potentials the last can correspond to subdiffusion. It was shown that the same behavior can be found in realistic polymer chain models. Therefore a one-dimensional chain with double-well potentials might be a possible minimal model describing the similar observations of anomalous kinetics in realistic molecular dynamics simulations. Moreover, the model is a close continuous analogue of the so-called ``necklace" model for reptation, so that the results might have a broader applicability also outside of the scope of present investigations.

\begin{acknowledgments}
The authors thankfully acknowledge financial support by DFG within the SFB 555 research collaboration program.
\end{acknowledgments}


\begin{thebibliography}{18}
\expandafter\ifx\csname natexlab\endcsname\relax\def\natexlab#1{#1}\fi
\expandafter\ifx\csname bibnamefont\endcsname\relax
  \def\bibnamefont#1{#1}\fi
\expandafter\ifx\csname bibfnamefont\endcsname\relax
  \def\bibfnamefont#1{#1}\fi
\expandafter\ifx\csname citenamefont\endcsname\relax
  \def\citenamefont#1{#1}\fi
\expandafter\ifx\csname url\endcsname\relax
  \def\url#1{\texttt{#1}}\fi
\expandafter\ifx\csname urlprefix\endcsname\relax\def\urlprefix{URL }\fi
\providecommand{\bibinfo}[2]{#2}
\providecommand{\eprint}[2][]{\url{#2}}

\bibitem[{\citenamefont{Granek and Klafter}(2005)}]{Granek05}
\bibinfo{author}{\bibfnamefont{R.}~\bibnamefont{Granek}} \bibnamefont{and}
  \bibinfo{author}{\bibfnamefont{J.}~\bibnamefont{Klafter}},
  \bibinfo{journal}{Phys. Rev. Lett.} \textbf{\bibinfo{volume}{95}},
  \bibinfo{pages}{098106} (\bibinfo{year}{2005}).

\bibitem[{\citenamefont{Kou and Xie}(2004)}]{Xie04}
\bibinfo{author}{\bibfnamefont{S.~C.} \bibnamefont{Kou}} \bibnamefont{and}
  \bibinfo{author}{\bibfnamefont{X.~S.} \bibnamefont{Xie}},
  \bibinfo{journal}{Phys. Rev. Lett.} \textbf{\bibinfo{volume}{93}},
  \bibinfo{pages}{180603} (\bibinfo{year}{2004}).

\bibitem[{\citenamefont{Togashi and Mikhailov}(2007)}]{Togashi07_PNAS}
\bibinfo{author}{\bibfnamefont{Y.}~\bibnamefont{Togashi}} \bibnamefont{and}
  \bibinfo{author}{\bibfnamefont{A.~S.} \bibnamefont{Mikhailov}},
  \bibinfo{journal}{Proc. Natl. Acad. Sci. U.S.A.}
  \textbf{\bibinfo{volume}{104}}, \bibinfo{pages}{8697–8702}
  (\bibinfo{year}{2007}).

\bibitem[{\citenamefont{Cressman et~al.}(2008)\citenamefont{Cressman, Togashi,
  Mikhailov, and Kapral}}]{Cressmann08_PRE}
\bibinfo{author}{\bibfnamefont{A.}~\bibnamefont{Cressman}},
  \bibinfo{author}{\bibfnamefont{Y.}~\bibnamefont{Togashi}},
  \bibinfo{author}{\bibfnamefont{A.~S.} \bibnamefont{Mikhailov}},
  \bibnamefont{and} \bibinfo{author}{\bibfnamefont{R.}~\bibnamefont{Kapral}},
  \bibinfo{journal}{Phys. Rev. E} \textbf{\bibinfo{volume}{77}},
  \bibinfo{eid}{050901} (\bibinfo{year}{2008}).

\bibitem[{\citenamefont{Tang and Lin}(2006)}]{Tang06_PRE}
\bibinfo{author}{\bibfnamefont{J.}~\bibnamefont{Tang}} \bibnamefont{and}
  \bibinfo{author}{\bibfnamefont{S.-H.} \bibnamefont{Lin}},
  \bibinfo{journal}{Phys. Rev. E} \textbf{\bibinfo{volume}{73}},
  \bibinfo{eid}{061108} (\bibinfo{year}{2006}).

\bibitem[{\citenamefont{Neusius et~al.}(2008)\citenamefont{Neusius, Daidone,
  Sokolov, and Smith}}]{Neusius08}
\bibinfo{author}{\bibfnamefont{T.}~\bibnamefont{Neusius}},
  \bibinfo{author}{\bibfnamefont{I.}~\bibnamefont{Daidone}},
  \bibinfo{author}{\bibfnamefont{I.~M.} \bibnamefont{Sokolov}},
  \bibnamefont{and} \bibinfo{author}{\bibfnamefont{J.~C.} \bibnamefont{Smith}},
  \bibinfo{journal}{Phys. Rev. Lett.} \textbf{\bibinfo{volume}{100}},
  \bibinfo{eid}{188103} (\bibinfo{year}{2008}).

\bibitem[{\citenamefont{Senet et~al.}(2008)\citenamefont{Senet, Maisuradze,
  Foulie, Delarue, and Scheraga}}]{Senet08}
\bibinfo{author}{\bibfnamefont{P.}~\bibnamefont{Senet}},
  \bibinfo{author}{\bibfnamefont{G.~G.} \bibnamefont{Maisuradze}},
  \bibinfo{author}{\bibfnamefont{C.}~\bibnamefont{Foulie}},
  \bibinfo{author}{\bibfnamefont{P.}~\bibnamefont{Delarue}}, \bibnamefont{and}
  \bibinfo{author}{\bibfnamefont{H.~A.} \bibnamefont{Scheraga}},
  \bibinfo{journal}{Proc. Natl. Acad. Sci. U.S.A.}
  \textbf{\bibinfo{volume}{105}}, \bibinfo{pages}{19708}
  (\bibinfo{year}{2008}).

\bibitem[{\citenamefont{Fixman}(1978)}]{Fixman78}
\bibinfo{author}{\bibfnamefont{M.}~\bibnamefont{Fixman}}, \bibinfo{journal}{J.
  Chem. Phys.} \textbf{\bibinfo{volume}{69}}, \bibinfo{pages}{1538}
  (\bibinfo{year}{1978}).

\bibitem[{\citenamefont{Ryckaert and Bellemans}(1975)}]{Ryckaert75}
\bibinfo{author}{\bibfnamefont{J.~P.} \bibnamefont{Ryckaert}} \bibnamefont{and}
  \bibinfo{author}{\bibfnamefont{A.}~\bibnamefont{Bellemans}},
  \bibinfo{journal}{Chem. Phys. Lett.} \textbf{\bibinfo{volume}{30}},
  \bibinfo{pages}{123 } (\bibinfo{year}{1975}), ISSN \bibinfo{issn}{0009-2614}.

\bibitem[{\citenamefont{Binder and Paul}(1997)}]{Binder97}
\bibinfo{author}{\bibfnamefont{K.}~\bibnamefont{Binder}} \bibnamefont{and}
  \bibinfo{author}{\bibfnamefont{W.}~\bibnamefont{Paul}},
  \bibinfo{journal}{Journal of Polymer Science: Part B}
  \textbf{\bibinfo{volume}{35}}, \bibinfo{pages}{1} (\bibinfo{year}{1997}).

\bibitem[{\citenamefont{Rigby and Roe}(1987)}]{Rigby87}
\bibinfo{author}{\bibfnamefont{D.}~\bibnamefont{Rigby}} \bibnamefont{and}
  \bibinfo{author}{\bibfnamefont{R.-J.} \bibnamefont{Roe}},
  \bibinfo{journal}{The Journal of Chemical Physics}
  \textbf{\bibinfo{volume}{87}}, \bibinfo{pages}{7285} (\bibinfo{year}{1987}).

\bibitem[{\citenamefont{Kitao and Go}(1999)}]{Kitao99}
\bibinfo{author}{\bibfnamefont{A.}~\bibnamefont{Kitao}} \bibnamefont{and}
  \bibinfo{author}{\bibfnamefont{N.}~\bibnamefont{Go}},
  \bibinfo{journal}{Current Opinion in Structural Biology}
  \textbf{\bibinfo{volume}{9}}, \bibinfo{pages}{164 } (\bibinfo{year}{1999}).

\bibitem[{\citenamefont{Guidoni et~al.}(2003)\citenamefont{Guidoni, M\'artin,
  and Aldao}}]{Guidoni03}
\bibinfo{author}{\bibfnamefont{S.~E.} \bibnamefont{Guidoni}},
  \bibinfo{author}{\bibfnamefont{H.~O.} \bibnamefont{M\'artin}},
  \bibnamefont{and} \bibinfo{author}{\bibfnamefont{C.~M.} \bibnamefont{Aldao}},
  \bibinfo{journal}{Phys. Rev. E} \textbf{\bibinfo{volume}{67}},
  \bibinfo{pages}{031804} (\bibinfo{year}{2003}).

\bibitem[{\citenamefont{Drzewi\'{n}ski and van Leeuwen}(2006)}]{Drz06}
\bibinfo{author}{\bibfnamefont{A.}~\bibnamefont{Drzewi\'{n}ski}}
  \bibnamefont{and} \bibinfo{author}{\bibfnamefont{J.~M.~J.} \bibnamefont{van
  Leeuwen}}, \bibinfo{journal}{Physical Review E (Statistical, Nonlinear, and
  Soft Matter Physics)} \textbf{\bibinfo{volume}{73}}, \bibinfo{eid}{051801}
  (pages~\bibinfo{numpages}{7}) (\bibinfo{year}{2006}).

\bibitem[{\citenamefont{Doi}(1996)}]{Doi96}
\bibinfo{author}{\bibfnamefont{M.}~\bibnamefont{Doi}},
  \emph{\bibinfo{title}{Introduction to Polymer Phyics}}
  (\bibinfo{publisher}{Oxford University Press, Oxford}, \bibinfo{year}{1996}).

\bibitem[{\citenamefont{Toda}(1989)}]{Toda89}
\bibinfo{author}{\bibfnamefont{M.}~\bibnamefont{Toda}},
  \emph{\bibinfo{title}{Theory of Nonlinear Lattices}}
  (\bibinfo{publisher}{Springer, Berlin}, \bibinfo{year}{1989}).

\bibitem[{\citenamefont{Tournier and Smith}(2003)}]{Tournier03_PRL}
\bibinfo{author}{\bibfnamefont{A.~L.} \bibnamefont{Tournier}} \bibnamefont{and}
  \bibinfo{author}{\bibfnamefont{J.~C.} \bibnamefont{Smith}},
  \bibinfo{journal}{Phys. Rev. Lett.} \textbf{\bibinfo{volume}{91}},
  \bibinfo{pages}{208106} (\bibinfo{year}{2003}).

\bibitem[{\citenamefont{de~Gennes}(1979)}]{deGennes79}
\bibinfo{author}{\bibfnamefont{P.-G.} \bibnamefont{de~Gennes}},
  \emph{\bibinfo{title}{Scaling Concepts in Polymer Physics}}
  (\bibinfo{publisher}{Cornell University Press, London},
  \bibinfo{year}{1979}).

\end{thebibliography}

\newpage
\section*{Figure captions}

\subsection*{Figure~\ref{fig:CMPCs}.}
$1-\phi_k(t)$ of some of the PCs. Panel (a): High temperature value. Panel (b): Intermediate temperature value for which strong internal friction is observed. The parameter values are given in \cite{Rigby87} and in the text.

\subsection*{Figure~\ref{fig:dwETE}.}
ACF of the end-to-end distance $R_{ete}$. Shown is  $1-\phi_{ete}(t)$. Panel (a): Single-well interaction potentials. The noise strengths is $k_BT/\gamma=5$. Panel (b): DW-potential. The barrier heights are given in the legend, $b=2$.

\subsection*{Figure~\ref{fig:DWPCs}.}
(Color online) $1-\phi_k(t)$ of the lowest PCs. Panel (a): Single-well potentials. The noise strength is the same as in Fig.~\ref{fig:dwETE}. Panel (b): DW-potential with the parameter values $a=1$, $b=2$, and $e=10$. Panel (c): DW-potential with barrier heights distributed as described in the text.

\newpage

\section*{Figures}
\newpage

\newpage

\newpage

\end{document}